\documentstyle[12pt,epsfig]{article}
\date{November 13, 1995}

\title{
\rightline{\small UL--NTZ 30/95}
A new possibility to monitor collisions of relativistic
heavy ions at LHC and RHIC }

\author{R.~Engel$^{1,}$\thanks{eng@tph200.physik.uni-leipzig.de} ,
  A.~Schiller$^{1,}$\thanks{schiller@tph200.physik.uni-leipzig.de} \
 and  V.G.~Serbo$^{1,2,}$\thanks{serbo@tph100.physik.uni-leipzig.de}  \\
{\it $^1$ Institut f\"ur Theoretische Physik, Universit\"at
Leipzig, F.R. Germany}  \\
{\it $^2$ Novosibirsk State University, Novosibirsk, Russia}}

\begin{document}
\maketitle

\begin{abstract}
We consider the radiation  of
 particles of one bunch in the collective field of
the oncoming bunch,  called coherent bremsstrahlung (CBS).
The main characteristics of CBS for  LHC  (in the
Pb--Pb mode) and for RHIC are calculated.
At LHC about $3.9 \cdot 10^8 dE_\gamma /
E_\gamma$ photons per second are expected for photon energies
$E_\gamma \stackrel{<}
{\sim} E_c= 93 $ eV.
It seems that CBS can be a potential tool for optimizing
collisions and for measuring beam parameters. The bunch
length $\sigma_z$ can be found  from the critical energy of the CBS
spectrum $E_c \propto 1/\sigma_z$;
the transverse bunch size $\sigma_\bot$ is related to the photon rate
$dN_\gamma \propto 1/{\sigma_\bot}^2$.
A specific dependence of $dN_\gamma$ on the impact parameter between the beams
allows for a fast control over the beam displacement.
\end{abstract}

{\it 1. Introduction} --
In this paper  a new type of radiation at colliders -- the
coherent bremsstrahlung (CBS)-- is considered.
The CBS is the radiation of one
bunch particles in the short collective electromagnetic field of
the oncoming bunch.

For definiteness, let us consider the photon emission by a single ion with a
charge $Z_1 e$ moving through a bunch of ions with the charge
$Z_2 e$.  The ordinary (incoherent) brems\-strah\-lung of ions
at colliders has a  relatively small cross section. On the other hand,
at small enough photon energies $E_\gamma$, the radiation due to
interaction of the $Z_1$ ion  with the
second bunch becomes coherent resulting in
large  number of emitted photons. This increase arises from
 an extra factor $N_2$ where $N_2$ is the second bunch
population. Therefore, the number of the emitted photons $dN_\gamma$
is already proportional to ${N_2}^2$
\begin{equation}
dN_{\gamma}\; \propto \;N_1\; {N_2}^2\; {dE_{\gamma}\over E_{\gamma}} \ .
\label{1}
\end{equation}

The CBS  occurs
when a coherence length $\sim 4 {\gamma_1}^2 \hbar c / E_\gamma$
becomes comparable or larger to the length of the second bunch $\sigma_{2z}$
(here $\gamma_1 = E_1/(m_1 c^2)$ is the
Lorentz factor of the $Z_1$ ion).
At photon energies
\begin{equation}
 E_\gamma \stackrel{<}{\sim}
E_c=4{{\gamma_1}^2 \hbar c \over \sigma_{2z}}
\label{2}
\end{equation}
the radiation arises from the interaction of the $Z_1$ ion  with the
second bunch which looks like a  ``particle'' with the
huge charge $Z_2 e N_2$.

A classical approach to CBS was given in \cite{Bassetti}, a quantum
treatment of CBS and  applications  to some working and
planned colliders can be found in \cite{Ginz}-\cite{Serbo}. Recently we have
developed a new simple and transparent method to  calculate this radiation
\cite{ESS}.
For the application of CBS to relativistic heavy ion colliders
we collect the needed  formulae from \cite{Ginzyaf} and \cite{ESS}.

The method of calculation  is valid if the ion deflection angle $\theta_d$ in
the field of the oncoming bunch is small compared with the typical
radiation angle $\theta_r \sim 1/ \gamma_1$.
The ratio of these angles  is easily estimated knowing
the electric  and magnetic
 fields of the second bunch which are approximately equal
in magnitude,
$| {\bf E}| \approx |{\bf B}|   \sim  2 Z_2eN_2 /
(\sigma_{2z} \sigma_{2\bot})$
(for LHC the effective field $|{\bf E}| + |{\bf B}|
\sim 0.1$T)
\begin{equation}
{\theta_d \over \theta _r }  \sim
\eta={Z_2\over Z_1}{r_1 N_2\over \sigma_{2\bot}}
\label{3}
\end{equation}
where $\sigma_{2\bot}$  is the transverse size of the
second bunch and $r_1 ={Z_1}^2 e^2/(m_1 c^2)$ is the classical ion radius.

For  LHC (in the mode Pb-Pb) and for  RHIC  the
parameter $\eta$ is of the order
\begin{equation}
\eta \sim 10^{-4}-10^{-3},
\label{4}
\end{equation}
so the above mentioned method is valid.

It is worthwhile to notice that
for the majority of the colliders  $\eta$ is either much smaller
than one (all the $pp$, $\bar p p$, some  $e^+e^-$
colliders and  B-factories) or $\eta \sim 1$ (e.g., LEP,
TRISTAN) and only the linear $e^+e^-$
colliders have $\eta \gg 1$. Therefore, the CBS has a very
wide region of applicability.

In calculating the CBS effect we  have   assumed
 that the densities of the bunches  do not
change during the collisions and that the bunches in the
interacting region have Gaussian particle distributions with
mean-squared transverse  $\sigma_{j\bot}$
and longitudinal $\sigma_{jz}$ radii, $
j=1,2$.
All parameters which are necessary for the calculation are given in
Table~\ref{tab1}, for LHC they are taken from \cite{PDG} and for RHIC from
\cite{Peggs}. For collisions of identical beams we assume $N_1=N_2=N$.
\begin{table}[htb]
\caption{\label{tab1} Parameters of some colliders}
\renewcommand{\arraystretch}{1.5}
\begin{center}
\par
 \begin{tabular}{|c|c|c|c|c|}\hline
& $N$ & $\gamma = E/( mc^2)$ & $\sigma_z $ (cm) & $\sigma_{\bot} \;
(\mu$m) \\ \hline
$Pb$ (LHC) & $0.9 \cdot 10^8$ & 2980 & 7.5 & 15  \\ \hline
$Au$ (RHIC) & $10^9$ & 108 & 11.9 & 150  \\ \hline
$p$ (RHIC)  & $10^{11}$ & 268 & 7.2 & 110  \\  \hline
\end{tabular}
\end{center}
\end{table}

{\it 2. Spectrum of CBS photons} ---
The number of CBS
photons for {\it a single collision} of the beams is equal to
\begin{equation}
dN_{\gamma }=N_{0}\,\Phi (E_{\gamma}/E_{c})\,{dE_{\gamma}\over
E_{\gamma}}.
\label{5}
\end{equation}
Here for the round Gaussian bunches with
\begin{equation}
\sigma _{1\bot} = \sigma _{1x} = \sigma_{1y}; \;\;\;
\sigma _{2\bot} = \sigma _{2x} = \sigma_{2y} \ ,
\label{6}
\end{equation}
the constant $N_0$ is equal to
\begin{equation}
N_0={4\over 3}\;{\alpha \over \pi}\; N_1\; \left({Z_2 r_1 N_2 \over
\sigma _{1\bot}}\right) ^2\;
\ln{{ (\sigma _{1\bot}^2 + \sigma _{2\bot}^2)^2\over
2\sigma _{1\bot}^2\,\sigma_{2\bot}^2+ \sigma_{2\bot}^4 }} \ .
\label{7}
\end{equation}
For identical beams
\begin{equation}
\sigma_{1\bot}=\sigma_{2\bot} = \sigma_\bot ,
\label{8}
\end{equation}
 the formula (\ref{7})  simplifies
\begin{equation}
N_0={4\over 3}\;\ln{{4\over 3 }}\;  {\alpha\over \pi}\;  N_1 \;
\left({Z_2 r_1 N_2 \over \sigma _\bot}\right) ^2 =
8.91 \cdot 10^{-4}\; N_1\left({Z_2 r_1 N_2 \over \sigma
_\bot}\right) ^2 \ .
\label{9}
\end{equation}
The function $\Phi(x)$ is defined as follows
\begin{equation}
\Phi (x)={3\over 2}\;\int _0^\infty {1+z^2\over (1+z)^4}\; \mbox
{exp} [-x^2(1+z)^2]\;dz;
\label{10}
\end{equation}
\begin{equation}
\Phi (x)=1 \;\;\mbox{at}\;\; x \ll 1;\;\;\; \Phi
(x)=(0.75/x^2)\cdot \mbox{e}^{-x^2} \;\;\mbox{at}\;\; x\gg 1
\label{11}
\end{equation}
(some values of this function are: $\Phi (x)=$ 0.80, 0.65, 0.36,
0.10, 0.0023 for $x=$ 0.1, 0.2, 0.5, 1, 2).

The calculated constants $N_0$, the critical energies
$E_c$ and  wave lengths $\lambda_c =
2\pi \hbar c / E_c$ are presented in Table~\ref{tab2}.
\begin{table}[htb]
\caption{\label{tab2} Parameters of CBS}
\renewcommand{\arraystretch}{1.5}
\begin{center}
\par
 \begin{tabular}{|c|c|c|c|c|}\hline
  &$Pb$-$Pb$ (LHC)&$Au$-$Au$ (RHIC)&$p$-$Au$ / $Au$-$p$ (RHIC)&$p$-$p$ (RHIC)
\\ \hline
$N_0 $                & 49            & 590            &
49 / 1800       & 170  \\ \hline
$E_c$ (eV)            & 93            & 0.077          &
0.48  / 0.13    & 0.79  \\  \hline
$\lambda_c $ ($\mu$m) & 0.013         & 16             &
2.6 / 9.7 & 1.6  \\ \hline
\end{tabular}
\end{center}
\end{table}

Note the following features of the CBS spectrum. The constant $N_0$
is proportional to $1/{\sigma_\bot}^2$ and the shape of the spectrum
strongly depends on the bunch length $\sigma_{2z}$ via Eqs. (\ref{2})
and (\ref{5}).
Therefore, measuring the photon rate and the shape of the spectrum
one can obtain informations about the beam sizes.

It may be convenient for LHC to use the CBS photons in the
range of {\it visible light} $E_\gamma \sim 2-3$ eV $\ll E_c=93$
eV.  In this region the rate of photons is expected to be
\begin{equation}
{dN_\gamma \over \tau} \approx 3.9 \cdot 10^{8}\; {dE_\gamma \over
E_\gamma} \;\; \;\mbox{photons$\;$ per $\;$ second}
\label{12}
\end{equation}
(here $\tau=0.125 \; \mu$s \cite{Eggert} is the time between bunch collisions
at a given interaction point). Furthermore,  for visible light the photon
polarization should be easily measurable.

Let us discuss the question of a possible
background due to synchrotron radiation (SR) on
the external magnetic field of the
collider $B_{\rm ext}$.
A complete treatment of
this background can be done only knowing the details of
the collider magnetic layout. Nevertheless, for all the  discussed
heavy ion colliders we expect this
background to be small.

A rough estimation of the SR rate can be performed as follows.
The first bunch moving through the uniform field $B_{\rm ext}$
emits $dN_\gamma^{SR}$ photons
in the time $\Delta t$
\begin{equation}
dN_\gamma^{SR}= N_1 \  {d I \over E_\gamma} \  \Delta t=
{4\over9} \  \alpha \  N_1 \ F(E_\gamma/E_c^{SR}) \ {d E_\gamma
\over E_\gamma} {{Z_1}^3 e B_{\rm ext} \over m_1 c} \ \Delta t \ .
\label{121}
\end{equation}
Here $dI$ is the classical synchrotron radiation intensity per unit time,
$E_c^{SR}$ the critical SR energy
\begin{equation}
E_c^{SR} =  \hbar \ {3  Z_1 e B_{\rm ext} { \gamma_1}^2 \over  2 m_1 c } =
             \hbar \ { 3 {\gamma_1}^3 c \over 2 R_B}
\label{122}
\end{equation}
with the circular orbit of radius $R_B$. The normalized function $F(y)$
is defined as an integral over the modified Bessel function
$K_{5/3}(x)$ \cite{PDG}
\begin{equation}
F(y)= {9 \over 8 \pi} \ \sqrt{3} \ y \ \int_y^{\infty} \ K_{5/3}(x) \ dx \ .
\label{123}
\end{equation}
Since the CBS radiation is confined to angles $\stackrel{<}{\sim} 1/ \gamma_1$.
Only those SR photons should be considered which are
emitted in the same angular interval. Using this condition one can estimate
$\Delta t$
\begin{equation}
\Delta t \sim {R_B \over \gamma_1 c} = {m_1 c \over Z_1 e B_{\rm ext}}
\label{124}
\end{equation}
which allows to write Eq.~(\ref{121}) in the form
\begin{equation}
dN_\gamma^{SR} \sim
{4\over9} \ \alpha \  N_1 \ {Z_1}^2 \ F(E_\gamma/E_c^{SR}) \ {d E_\gamma
\over E_\gamma} \ .
\label{125}
\end{equation}

\begin{figure}[htb]
\epsfig{file=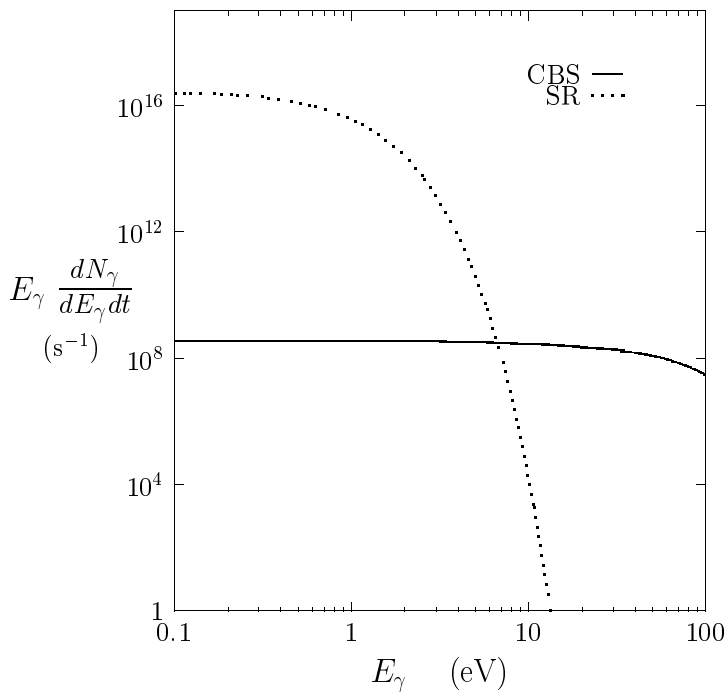,width=10cm}
\caption{\em
 Number of photons emitted due to CBS and SR ($B_{\rm ext} = 1$T)
at LHC (Pb-Pb) as function of the photon energy.
\label{lhc-cbs} }
\end{figure}

In Fig.~\ref{lhc-cbs} we compare the rate for the CBS and SR photons using
Eqs.~(\ref{5}), (\ref{125}) and the parameters of LHC assuming
an external field $B_{\rm ext} = 1$T, for definiteness. The critical SR
energy for this field strength is 0.35 eV (remember $E_c = 93 $ eV for CBS).
The ratio of the CBS rate to the SR rate changes drastically with the
photon energy. It is equal to 1, 10, and 1000 for
$E_\gamma= 6.5, 7.5$, and $9.8$ eV.
So we conclude that for photon energies  above $8$eV the synchrotron
radiation can be neglected.
It may be useful to consider the case $E_\gamma = E_c \gg E_c^{SR}$
where the signal to background ratio can be estimated as follows
\begin{equation}
{dN_\gamma \over dN_\gamma^{SR}} \sim 0.03
\left({ Z_2 r_1 N_2 \over Z_1 \sigma_\bot}\right)^2 \ { \mbox{e}^x\over
\sqrt{x}}  \ , \ \ \ x={E_c \over E_c^{SR}} \ .
\label{126}
\end{equation}

Let us stress that the CBS spectrum is completely different to that
of the synchrotron radiation in an uniform magnetic field.
Both are classical radiations in the considered cases, but CBS is an emission
in a short electromagnetic filed at the length scale of the bunch while
SR corresponds to photon emission in a static magnetic filed with a much
larger characteristic length scale.
In particular, the critical energy  $E_c$ of CBS (\ref{2})
does not depend neither on the charge nor on
the mass of the radiating ion contrary to
$E_c^{SR}$.

\begin{figure}[htb]
\epsfig{file=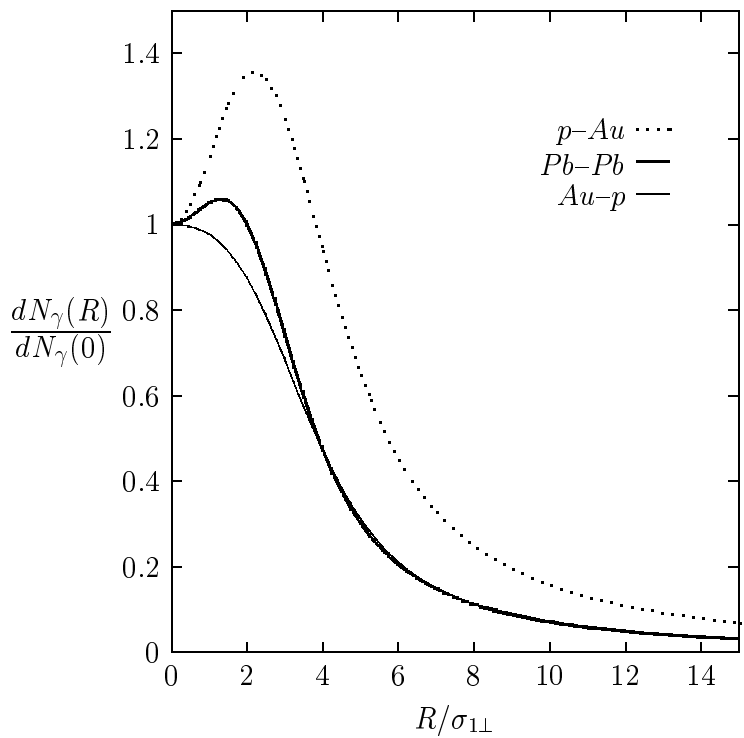,width=10cm}
\caption{\em Normalized number of CBS photons as function of the
impact parameter $R$.
\label{ion1} }
\end{figure}

If the first bunch  axis  is  transversely shifted
by a distance $R$ from the second bunch axis,
the luminosity $L(R)$ (as well as the number of events  for
the  ordinary reactions) decreases very quickly
\begin{equation}
L(R)=L(0) \, \exp \left(-{R^{2}\over 2(\sigma_{1\bot}^2+
\sigma_{2\bot}^2) }\right) \ .
\label{13}
\end{equation}
On the contrary, for the discussed colliders
the number of CBS photons decreases slowly and for small values
of $R$ this number even increases.
For the collision of identical round beams the
maximum of the ratio $dN_\gamma (R) /dN_\gamma(0)$ is equal to  $1.06$ at $R
\approx 1.3 \,\sigma _\bot$ (see Fig.~\ref{ion1})
whereas  the luminosity already decreases significantly.
At large displacements $ R \gg \sigma_\bot  $ this ratio is
\begin{equation}
{dN_\gamma (R) \over dN_\gamma(0)} = 6.95 \ { {\sigma_\bot}^2 \over R^2} \ .
\label{131}
\end{equation}
Additionally, two cases of non-identical beams
(p-Au and Au-p collisions at RHIC) are also
shown in Fig.~\ref{ion1}.
The dependence of CBS on the beam axis displacement can be used for a
fast control over the
collision impact parameter between beams at the interaction point
(especially during the beginning
of every run) and over the transverse beam size.  For the case
of another type of radiation (beamstrahlung), such an
experiment has already been successfully
performed at SLC \cite{Bon}.

\begin{figure}[htb]
\epsfig{file=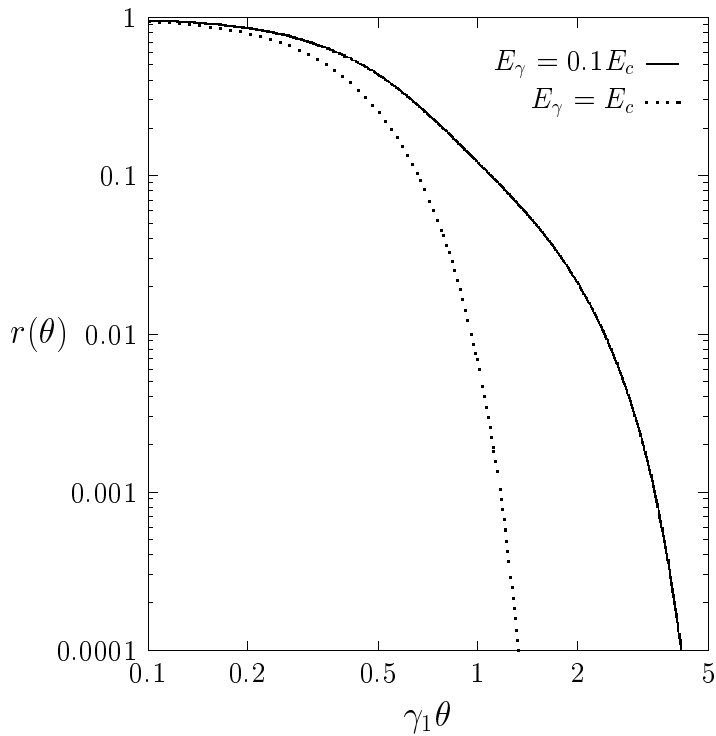,width=10cm}
\caption{\em
Angular distribution of the CBS photons for two typical energies,
normalized at $\theta = 0$,
$r(\theta) = dN_{\gamma}(\theta,E_\gamma)/dN_{\gamma}(0,E_\gamma)$.
\label{lhc-ang}}
\end{figure}

{\it 3. Azimuthal asymmetry and polarization} ---
The angular distribution for central head-on collisions of round
beams  has no azimuthal asymmetry, it is given by the expression
\begin{equation}
dN_\gamma = {3\over 2} N_0 \ {dE_\gamma \over E_\gamma} \
{1+z^2 \over(1+z)^4} \ dz  \
\mbox{exp} \left[ - \left( (1+z) {E_\gamma \over E_c}\right)^2 \right]
\label{132}
\end{equation}
where $ z = (\gamma_1 \theta)^2$, $\theta$ is  the polar angle
of the emitted photons.
In Fig.~\ref{lhc-ang} we present the normalized
angular distribution for different photon energies as function of
$\gamma_1 \theta$.
For small photon energies ($E_\gamma \ll E_c$) it
almost coincides with that of the ordinary bremsstrahlung. At photon energies
of the order of $E_c$ the distribution shrinks considerably.

It may be useful to note that in the energy region
\begin{equation}
  E_c^{SR} \ll E_\gamma \ll E_c
\label{1321}
\end{equation}
the angular distributions of CBS and SR photons are quite different. According
to (\ref{132}) the
characteristic emission angle $\theta_{\rm char}$ of CBS in this region is
\begin{equation}
  \theta_{\rm char} \approx {1 \over \gamma_1} \ .
\end{equation}
For photon energies (\ref{1321}) the distribution of SR photons over
 an angle $\vartheta$ (defined as the angle between
the photon momentum and the orbit plane) can be obtained from
\cite{LL} (see \S 74) as follows
\begin{equation}
{d N_\gamma^{SR} \over d E_\gamma d z^{SR} } \propto { 1 + 2 z^{SR} \over
\sqrt{ 1 + z^{SR}} } \ \mbox{ exp} \left( - {E_\gamma \over E_c^{SR} } \
(1+z^{SR} )^{3/2} \right), \ \ \ z^{SR}=(\gamma_1 \vartheta)^2 \ .
\end{equation} From
this expression one can see that the SR photon rate is strongly suppressed
for angles $\vartheta$ larger than
\begin{equation}
 \vartheta_{\rm char} \approx {1 \over \gamma_1} \sqrt{ {2 E_c^{SR} \over
 3 E_\gamma}} \ .
\end{equation}
The characteristic angle of SR is considerably smaller than that
of CBS. This fact can be used to suppress the SR background even stronger.

\begin{figure}[htb]
\epsfig{file=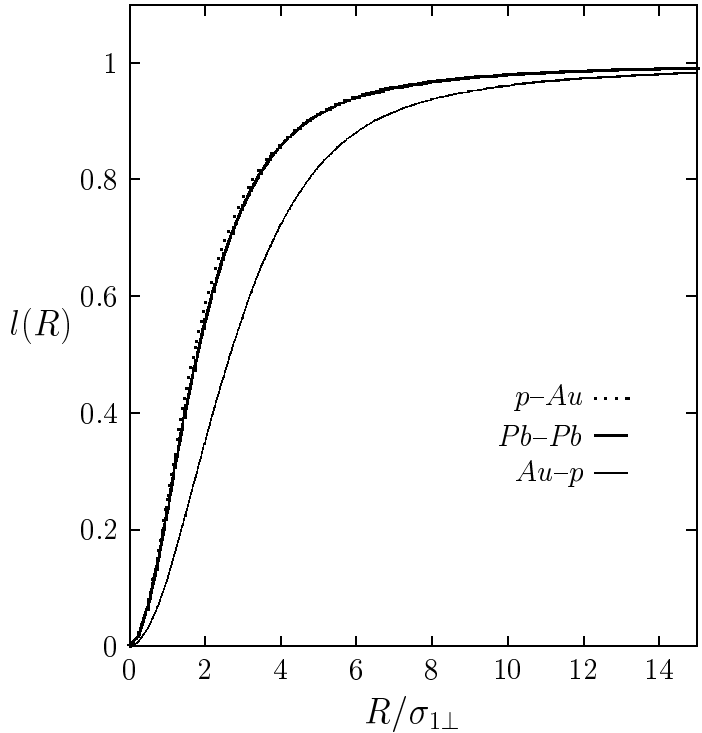,width=10cm}
\caption{\em
Average degree of linear polarization as function of the
impact parameter $R$.
\label{ion2} }
\end{figure}

If the impact parameter $R$ between beams is non-zero,
the angular distribution over $\theta$ (or $z$) is unchanged,
but an azimuthal asymmetry of the CBS photons appears which can also be used
for an operative beam control.
For example, if $R$ increases in the vertical direction, the first
bunch is shifted to the region where the electric field of
the second bunch is directed almost vertically.
This moving field of the second bunch
can be represented as a flux of equivalent
photons \cite{ESS}.
These photons are linearly  polarized in the vertical direction. The average
degree of such a polarization $l$ for the  discussed colliders with
identical and non-identical beams as function of $R$ is shown in
Fig.~\ref{ion2}.

Let us define the azimuthal asymmetry of the emitted photons
by the relation
\begin{equation}
A= {dN_{\gamma}(\varphi=0)-dN_{\gamma}(\varphi=\pi/2)\over
dN_{\gamma}(\varphi=0)+dN_{\gamma}(\varphi=\pi/2)}\ ,
\label{14}
\end{equation}
where the azimuthal angle $\varphi$ is measured with respect to
the horizontal plane. This
quantity does not depend on the photon energy and is equal to
\begin{equation}
A={2(\gamma_1 \theta)^2 \over 1+(\gamma_1 \theta)^4} \; l(R) \ .
\label{15}
\end{equation}
Increasing  $R$ in the vertical direction (see Fig.~\ref{ion2}),
the fraction of
photons emitted horizontally becomes larger than
the fraction of photons emitted vertically.

Finally, let us briefly discuss the polarization of CBS photons.
For linearly polarized equivalent photons
the CBS photons also get a
linear polarization in the same direction. Denoting by $l^{CBS}$
the average degree of CBS photon polarization, the ratio
$l^{CBS}/l$ varies as function of $E_\gamma$ in the interval
from 0.5 to 1 \cite{Serbo} (see Table~\ref{tab3}).
\begin{table}[htb]
\caption{\label{tab3} Degree of the linear polarization of the
CBS photons}
\renewcommand{\arraystretch}{1.5}
\begin{center}
\par
 \begin{tabular}{|c|c|c|c|c|c|c|c|c|}\hline
 $E_\gamma /E_c$ & 0 & 0.2 & 0.4 & 0.6 & 0.8 & 1 & 1.5 & 2
 \\ \hline $l^{CBS}/l$
& 0.5 & 0.7 & 0.81 & 0.86 & 0.89 & 0.94 & 0.96 & 0.97  \\
\hline
\end{tabular}
\end{center}
\end{table}

{\it Conclusions ---}\\
We have calculated the main characteristics  of CBS photons for
heavy ion colliders,  the photon rate, the energy and angular
distributions as well as the polarization.
It seems that CBS can be a potential tool for optimizing
collisions and for measuring beam parameters directly at the interaction
point. Obtaining the critical energy $E_c$ from the CBS spectrum,
the bunch length $\sigma_z$ can be found since
$E_c$ is proportional to $1/\sigma_{z}$. The transverse bunch size
$\sigma_\bot$ is related to the rate of photons
$dN_\gamma \propto 1/{\sigma_\bot}^2$. The possible background due to
synchrotron radiation on the external magnetic field of the collider is
estimated.
Furthermore, CBS may be very
useful for a fast control over the impact parameter
between the colliding bunch axes  because the photon rate
$dN_\gamma$ depends on this parameter in a very specific way.

{\it Acknowledgements} ---
We are grateful to K. Eggert for a useful discussion
and S. Peggs for providing us the RHIC parameters.
V.G.~Serbo acknowledges  support of the S\"achsisches Staatsministerium
f\"ur Wissenschaft und Kunst, of the
Naturwissenschaftlich--Theoretisches Zentrum of the Leipzig University
and of the Russian Fond of Fundamental Research.
R.~Engel is supported by the Deutsche
Forschungsgemeinschaft under grant Schi 422/1-2.

\end{document}